\documentclass[aps,pra,reprint,showpacs]{revtex4-1}
\usepackage[T1]{fontenc}
\usepackage[latin9]{inputenc}
\usepackage{amstext,amsmath}
\usepackage{amssymb}
\usepackage{graphicx}
\usepackage{color}
\usepackage{braket}
\usepackage{dsfont}

\newcommand{\tr}{\mathrm{tr}}

\begin{document}

\title{Multi-photon absorption in optical gratings for matter waves}

\author{Kai Walter$^1$, Stefan Nimmrichter$^2$, and Klaus Hornberger$^1$}

\affiliation{$^{1}$University of Duisburg-Essen, Faculty of Physics, Lotharstra{\ss}e
1-21, 47048 Duisburg, Germany\\
$^2$Centre for Quantum Technologies, National University of Singapore, 3 Science Drive 2,
Singapore 117543}


\begin{abstract}

We present a theory for the diffraction of large molecules or nanoparticles at a standing light 
wave. Such particles can act as a genuine photon absorbers
due to their numerous internal degrees of freedom effecting fast internal energy conversion. Our 
theory incorporates the interplay
of three light-induced properties: the coherent phase modulation due to the dipole interaction,
a non-unitary absorption-induced amplitude modulation described as a generalized measurement, and 
a coherent recoil splitting that resembles a quantum random walk in steps of the photon momentum. 
We discuss how these effects show up in near-field and far-field interference schemes, and we
confirm our effective description by a dynamic evaluation of the grating interaction, which 
accounts for the internal states.
\end{abstract}

\maketitle

\section{Introduction}
Exploring matter-wave interference with heavy molecules and nanoparticles is of fundamental 
interest, as it allows exploring the possible limits of quantum mechanics at macroscopic scales
\cite{RevModPhysBassi,hornberger2012colloquium}, and developing new 
tools to measure accurately internal molecular properties \cite{Eibenberger2014}.

Motivated by experiments 
\cite{abfalterer1997,arndt1999wave,PhysRevLett.87.160401,andrey2003,C3CP51500A,Eibenberger2014,Cotter2015} that use a standing 
light field
as diffraction element, we focus on the interaction between a
standing-wave laser grating and a delocalized and internally complex molecule.
It has been seen earlier that spontaneous emission, light scattering, or thermal 
radiation induced by an optical grating can lead to decoherence 
\cite{PhysRevLett.75.3783,NatureScully,PhysRevA.71.023601}.
However, complex molecules tend to absorb the light from the laser grating predominantly without 
subsequent reemission because they can rapidly redistribute the photon energy to many internal 
states, acting effectively as an energy sink. This applies all the more so to large, optically 
levitated nanoparticles, e.g. silica spheres, which are candidates for novel optomechanics and 
matter-wave interference schemes \cite{chang2010cavity, romero2010toward, romero2011large, 
NatComBateman}.
It is intriguing to ask whether such molecules are capable to interfere
after the absorption given that it might
reveal ``which-way'' information and the position of the molecular center of
mass. In the following we discuss how the molecules are still
able to interfere even after absorption of many laser photons. 

Specifically, the theory developed below is required to quantitatively describe a recent experiment 
with C$_{70}$ fullerenes which featured a high laser power and unprecedented velocity resolution 
\cite{Cotter2015}. It provides 
evidence that the momentum recoil upon photon absorption from a standing light field is coherent in 
the sense that it leads to a superposition of the momentum kicks associated with the two 
possible photon directions, rather than to a mixture. Despite the state-insensitive detection and 
hence the lack of coherence in the number of absorptions, the experiment rules out a classical 
random-walk description for the absorption process.

The structure of this article is as follows. 
In Sect. \ref{sec:optical-grating} we make use of the formalism of generalized measurements  
\cite{NC,peres, 
breuer2002theory, wiseman2009quantum, MichelQuantum} to describe 
all relevant aspects of the light-particle interaction: the influence of the dipole force, the 
stochastic impact of photon absorptions, and the matter-wave amplitude modulation associated 
with the absorption-induced postselection. 
The influence of these effects on matter-wave interferometry is then discussed in 
Sect.~\ref{sec:interference} by considering near-field and far-field schemes. 
In the near-field case, photon absorption modulates the periodic fringe pattern and lowers the 
interference visibility on average, while in the far-field case, photon absorption leads to new 
features in the observed interferogram. The nonclassical nature of the absorption recoil is only 
visible in the near field.
In Sect.~\ref{sec:master} we develop dynamic models of the laser-grating interaction in order 
to corroborate and generalize the measurement-based description of Sect.~\ref{sec:optical-grating}. 
A ladder model for the internal state allows evaluating the center-of-mass quantum dynamics 
in presence of absorption, accounting for possible photo-induced changes of the optical molecular 
properties. Moreover, we consider a three-level model with a dark state in order to
incorporate partially coherent Rabi-oscillations expected for resonant transitions.
We present our conclusions in Sect.~V.

\section{\label{sec:optical-grating}Optical grating transformation as a generalized measurement}

We start by developing an effective measurement-based description of the interaction
between a standing-wave laser grating and an absorbing molecule in terms of a generalized 
measurement transformation. It serves to incorporate
the essential effect of photon absorption on the center-of-mass motion of complex molecules in the 
absence of detailed knowledge about the molecule's level structure and transition dipole moments.
Avoiding a microscopic treatment, we model the interaction
effectively by means of the particle's complex susceptibility including the dipole 
polarizability and the absorption cross section at a given laser wavelength. 
Our approximate model would fail if the life times of the electrically excited states were 
long compared with the interaction time, see Sect.~\ref{sec:rabi}. For large molecules the 
relaxation time is typically on the order of picoseconds so that our effective description should 
be valid. For even larger nanoparticles, the optical response is often entirely characterized by a 
phenomenological dielectric function, which may exhibit broad internal (plasmonic) resonances at 
optical-to-UV wavelengths \cite{mayergoyz2013plasmon}.   

The model for the scattering of the molecular center-of-mass motion off the 
standing light field comprises two independent steps: 
On the one hand, a unitary state transformation  describes 
the coherent matter-wave phase modulation due to the dipole interaction, see 
Sect.~\ref{sec:phasemod}. On the other hand, the non-unitary transformation developed in 
Sect.~\ref{sec:Photon-absorption} accounts for the change of state due to photon 
absorption. In the context of interferometry,  it is convenient to represent the 
motional quantum state of the molecular center of mass in phase space. 
We therefore formulate the grating transformation in terms of the 
Wigner function \cite{Schleich} in Sect.~\ref{sec:phase_space}. The present formalism applies to 
all experimental scenarios where nanoparticles interact with optical standing waves in the 
Raman-Nath, or short-time, regime.

\subsection{Phase modulation} \label{sec:phasemod}
A sub-wavelength molecule or nanoparticle interacts with off-resonant light fields mainly through 
its frequency-dependent dipole polarizability $\alpha_{\mathrm{SI}}$. The particle is then subject 
to the dipole force proportional to the gradient of the local time-averaged standing-wave light 
intensity, 
\begin{equation}
I(x,y,z)=\frac{8P}{\pi w_{y} w_{z}}\exp\left(-\frac{2y^{2}}{ w_{y}^2}-\frac{2z^{2}}{ 
w_{z}^2}\right)\cos^{2}\left(k_{\mathrm{L}}x\right).\label{eq:intensity}
\end{equation}
Here, $x$ denotes the standing-wave axis, while $k_{\mathrm{L}}$, $w_y$, $w_z$ and $P$ are the wave number, the waists, and the running-wave power of a retro-reflected gaussian laser beam forming the standing wave.
In the Raman-Nath regime \cite{Moharam198019} of a short interaction time and high kinetic energy,
when the molecule rapidly crosses the laser beam at an approximately constant velocity $v_z$ in the
$xz$-plane, the dipole interaction is captured by the time-dependent potential
$V(x,t)=-(2\pi\alpha_\mathrm{SI}/4\pi \varepsilon_0c) I(x,0,v_{z}t)$. If also the transverse motion
along $x$ and $y$ can be neglected during the passage, the scattering problem effectively reduces to
one dimension. The molecule acquires an $x$-dependent phase
\cite{PhysRevA.78.023612,KDTLI} that results  in a unitary scattering transformation,
\begin{equation}
\mathsf{U}=\exp\left[  i \phi(\mathsf{x})\right] = \int d x \, \exp\left[  i \phi(x)\right]
|x\rangle\langle x|, \label{eq:unitary}
\end{equation}
with $\mathsf{x}$ the one-dimensional position operator in the standing-wave direction. This 
position-dependent phase shift $\phi(x)$ is given by the eikonal action accumulated during passage, 
i.e.~by the time integral of the interaction potential,
\begin{eqnarray}
\phi(x)&=&-\frac{1}{\hbar}\int_{-\infty}^{\infty} d t\:
V(x,t)=\phi_{0}\cos^{2}\left(k_{\mathrm{L}}x\right),\\
\phi_{0}&=&\frac{2\sqrt{2}}{\sqrt{\pi}\varepsilon_0}\frac{\alpha_\mathrm{SI}}{\hbar c}\frac{P}{
w_{y}v_{z}}.\label{eq:phi}
\end{eqnarray}
For realistic sub-wavelength molecules or nanoparticles, corrections  to
this  simple scattering model can be attributed to their anisotropy and to photon absorption. 
We focus on the latter effect in the following. For highly
anisotropic                                                                                   
 molecules, described in terms of a polarizability tensor, the fast molecular rotations
will lead to phase averaging and thus to degraded interference \cite{Stickler2015}.

\subsection{Photon absorption\label{sec:Photon-absorption}}

Standing light waves have often been employed as pure phase gratings
\cite{Freimund2001,Gould1986,PhysRevLett.75.2633,McGowan1995,NatKDTLI,PhysRevLett.87.160401, 
Chan1997, Martin1988}.
In the case of atoms and small molecules, the absorptionless phase-grating regime is 
achieved by detuning the laser wavelength sufficiently from internal resonances. 
The treatment of 
large molecules and nanoparticles requires a different approach since the
numerous rovibrational degrees of freedom give rise to a landscape of broad (collective) resonances 
in the
absorption spectrum $\sigma_{\mathrm{abs}}(\omega)$. The linear response of the molecule to the 
field is
described by a complex susceptibility that accounts for both the dipole interaction and the
absorption, $\chi = \alpha_\mathrm{SI} +  i c \varepsilon_0
\sigma_{\mathrm{abs}}/\omega_{\mathrm{L}}$ \cite{hulst1957light}. Photon absorption is negligible as 
long as $\beta =
\mathrm{Im} (\chi) / \mathrm{Re} (\chi) \ll 1$.

For lack of a microscopic description, we base our absorption model solely on the 
knowledge of collective properties that can be measured independently, such as the 
absorption cross-section $\sigma_{\mathrm{abs}}$ and the heat capacity $C$, as well as on the 
assumption of an internal heat sink: The excess energy $\hbar \omega_{\mathrm{L}}$ of an absorbed 
photon is assumed to be ``dissipated'' immediately, i.e.~redistributed among the many internal 
degrees of freedom. This results in a mean increase of the internal microcanonical temperature by 
$\Delta T = \hbar \omega_{\mathrm{L}}/C$. 
In practice, this turns out to be an excellent approximation for many complex molecules and 
nanoparticles with $C/k_\mathrm{B}\gg 1$, as long as only few photons are absorbed and the particle does not heat up 
too much. In specific cases, it might be necessary to take a radiative reemission of internal excess 
energy into account, either by fluorescence or by thermal black-body radiation at high temperatures. 
This would result in additional decoherence, the theoretical description of which can be found 
elsewhere \cite{PhysRevA.70.053608,PhysRevA.71.023601}.

From the point of view of operational quantum mechanics \cite{busch2009operational}, photon 
absorption can be described as a generalized measurement transformation \cite{NC,peres, 
breuer2002theory, wiseman2009quantum, MichelQuantum}, as specified by a set of measurement operators 
$\{ \mathsf{M}_{\ell} \}$ 
with $\sum_{\ell=0}^{\infty}\mathsf{M}_{\ell}\mathsf{M}_{\ell}^{\dagger}=\mathds{1}$. In this 
framework, the number $\ell=0,1,2,\ldots$ of absorbed photons, as recorded by 
the excess energy of the internal degrees of freedom corresponds to the measurement result. 
In principle, this thermal encoding 
can be revealed by a calorimetric absorption detection scheme, which yields the number $\ell$ of absorbed
photons by measuring their internal energy.
When the measurement indicates 
the absorption of $\ell$ photons the conditional transformation of the reduced 
center-of-mass state $\rho$ reads
\begin{equation}
\rho\rightarrow\rho_{\ell} := 
\frac{\mathsf{M}_{\ell}\rho\mathsf{M}_{\ell}^{\dagger}}{P_{\ell}(\rho)}, \qquad P_{\ell} (\rho) = 
\tr\left[\mathsf{M_{\ell}^{\dagger}M_{\ell}\rho}\right]. \label{eq:state_trafo}
\end{equation}
Considering the particle-laser interaction in the Raman-Nath regime, i.e.~in the 
limit of the short interaction time, 
the transverse motion between two subsequent absorption events can be 
neglected. This implies that the measurement operators $\mathsf{M}_\ell$ are 
diagonal in position, which ensures that the absorption probability depends 
only on the spatial probability distribution, $P_\ell (\rho) = \int  d 
x \, p_\ell(x) \langle x |\rho|x \rangle$. Here, $p_\ell(x)=\vert 
M_\ell (x)\vert^2$ represents the position dependent absorption 
probability. The measurement operator is therefore of the form
\begin{equation}
 \mathsf{M}_{\ell} = \int  d x \, M_{\ell} (x) |x
\rangle\langle x| =\mathsf{U}_\ell \left\vert \mathsf{M}_\ell(\mathsf{x}) \right\vert 
 \label{eq:Ml}
\end{equation} 
with a unitary, position-dependent phase $\mathsf{U}_{\ell} = \exp \left[ 
 i \phi_{\ell}(\mathsf{x}) \right]$, which does not change the absorption 
probability $p_\ell(x)$. The choice of the unitary is not 
arbitrary; it must be consistent with the
coherent phase modulation mediated by the molecular polarizability, as discussed in
the previous section.
Moreover, since the standing-wave field can be seen as a superposition of two
counter-propagating plane wave modes with wave vectors 
$k_\mathrm{L}\mathbf{e}_x$ and $-k_\mathrm{L}\mathbf{e}_x$, we expect that a 
superposition of recoils with momenta $\hbar k_\mathrm{L}$ and $-\hbar 
k_\mathrm{L}$ is coherently transferred to the particle upon absorption of a 
standing-wave photon. In particular, a relative phase appears between positions of a half-wavelength 
distances. If one photon is absorbed the diagonal elements must 
be proportional to $M_1(x)\propto \cos(k_\mathrm{L}x)$ because the momentum 
transfer of $\pm \hbar k_\mathrm{L}$ is described by the operator $\exp(\pm 
ik_\mathrm{L}\mathsf{x})$. 
Thus for $\ell$ subsequent photon absorptions the diagonal elements of the measurement 
operator are $M_\ell(x)\propto \cos^\ell(k_\mathrm{L}x)$, the zeros taking account 
for the fact that absorption does not take place at the nodes of the standing wave.

As a second step, we assume that the absorption cross-section and the 
polarizability do not change appreciably
upon absorption, i.e.~that the grating interaction does not depend on the internal state of the
molecule or nanoparticle. As discussed in Sect.~\ref{sec:master} this is a 
reasonable approximation for low laser powers; a generalization to the 
case of state-dependent internal properties is also presented in 
Sect.~\ref{sec:master}. The assumption of state independence implies that the 
number $\ell$ of absorbed photons follows a Poisson distribution, 
$ p_{\ell} (x) =  e^{-n(x)} n^\ell(x)/\ell !$,
with a mean value of
\begin{eqnarray}
n(x)&=&\int_{-\infty}^{\infty} d t\:\frac{\sigma_{\mathrm{abs}}}{\hbar
\omega_{\mathrm{L}}}I(x,0,v_{z}t)=n_{0}\cos^{2}(k_\mathrm{L}x), \\
n_{0}&=&\frac{8}{\sqrt{2\pi}}\frac{\sigma_{\mathrm{abs}}}{\hbar \omega_{\mathrm{L}}}\frac{P}{
w_{y}v_{z}}.\label{eq:n}
\end{eqnarray}
For the measurement operators (\ref{eq:Ml}) this means that the unitary part is 
independent of $\ell$, $\phi_{\ell} (x) = \phi(x)$, as given by the dipole 
interaction term (\ref{eq:phi}), and 
\begin{equation}
M_{\ell}(x)=\sqrt{\frac{n_{0}^{\ell}}{\ell!}}\cos^{\ell}\left(k_{\mathrm{L}}x\right) 
\exp\left(i\phi(x)-\frac{n(x)}{2}\right).\label{eq:measurment}
\end{equation}
The conditional state transformation (\ref{eq:state_trafo}) then describes the 
diffraction of particles after the absorption of  $\ell$ photons in the 
standing-wave grating.
In principle, these particles could be postselected in a 
detector sensitive to the internal energy. When the detector is 
insensitive to the internal state, one must average over all possible absorption 
numbers and apply the unconditional Kraus map \cite{MichelQuantum} 
\begin{equation}
\rho \to \rho' = \sum_{\ell=0}^{\infty}P_{\ell}(\rho)\rho_{\ell}=\sum_{\ell=0}^{\infty}\mathsf{M}_{\ell}\rho\mathsf{M}_{\ell}^{\dagger}.\label{eq:uncond_rho}
\end{equation}
As we will see in Sect.~\ref{sec:master}, this transformation is consistent with the solution of a dynamical model for photon absorption. 

In practice, the Poisson absorption model yields instructive, mostly analytic 
results for the relevant states of the incident particles, 
as well as for the final interferograms in both the near field and the far 
field.
Yet, it is known from molecular spectroscopy that the absorption 
cross section can grow after photon absorption and intersystem 
crossing \cite{Haufler,Henari1992144}. The excited-state polarizability might differ as well, and 
we will incorporate these effects in Sect.~\ref{sec:effect_of_excited}. It turns 
out that such
complications play only a minor quantitative role in the regime of low average
absorption, $n_0 \lesssim 2$.

In order to illustrate the influence of absorption on matter-wave diffraction, 
let us examine how
the measurement operators act on a momentum eigenstate $|p\rangle$ of the 
molecule, i.e.~when a coherent plane matter wave hits the grating. The 
phase modulation (\ref{eq:unitary})-(\ref{eq:phi}) leads to diffraction peaks separated by the 
grating momentum 
$2\hbar k_{\mathrm{L}}$ \cite{Cotter2015},
\begin{equation}
 \mathsf{U}|p\rangle= e^{ i
\phi_{0}/2}\sum_{\nu=-\infty}^{\infty} I_{\nu} \left( i\frac{\phi_{0}}{2}\right)\ket{p+2\nu\hbar
k_{\mathrm{L}}}.
\end{equation} 
which follows from a Fourier decomposition of (\ref{eq:unitary}). The Fourier components are the
modified Bessel functions $I_\nu(x)=i^{-\nu}J_\nu(ix)$. This constitutes the ideal
phase-grating effect for transparent particles with $\sigma_{\mathrm{abs}} = 0$. 
If however,
$\sigma_{\mathrm{abs}}>0$, but no photon is absorbed, we must apply the 
measurement operator
\begin{align}
\mathsf{M}_{0} (\mathsf{x})\ket{p} &= e^{i\phi_0/2-n_0/4}
\sum_{\nu=-\infty}^{\infty}
I_{\nu}( i\phi_0/2 - n_0/4) \nonumber \\ 
& \times \ket{p+2\nu\hbar k_{\mathrm{L}}},
\end{align}
and a different interferogram would be observed. Note that the particle then gets diffracted even 
for $\phi_0=0$.
This additional source of diffraction is related to the conditional modulation 
of the matter-wave \textit{amplitude}:
The spatial density of the post-measurement state is 
redistributed towards the standing-wave nodes where it is more likely that 
no absorption took place. 
This conditional transformation is used to describe diffraction at optical 
depletion gratings
\cite{OTIMA, NatOtima, PhysRevLettNadine, dis_stef}, where only those particles arrive at the 
detector that have not absorbed any photon.

In the case of $\ell$ subsequent absorption processes, the conditional transformation is given by 
\begin{eqnarray}
\mathsf{M}_{\ell}
(\mathsf{x})|p\rangle&=&\frac{e^{i\phi_0/2 - 
n_0/4}}{2^{\ell}}\sqrt{\frac{n_{0}^{\ell}}{\ell!}}\sum_{
\nu=-\infty}^{\infty}I_{\nu} ( i\phi_0/2 - n_0/4) \nonumber \\
 &\times& \sum_{n=0}^{\ell}\binom{\ell}{n }
\ket { p+2\hbar k_{\mathrm{L}}\nu+(\ell-2n)\hbar k_{\mathrm{L}}},
\end{eqnarray}
as follows from Eq.~(\ref{eq:measurment}).
Apart from the conditional diffraction by amplitude modulation, the binomial sum accounts for the 
coherent transfer of photon recoils in units of $\hbar k_{\mathrm{L}}$. With each absorption event 
the momentum state splits coherently into two branches shifted by $\pm \hbar k_{\mathrm{L}}$, a 
particular quantum analogue of a Galton board. Note however that the present model differs from the 
well-known quantum random-walk realizations of a Galton board found in the literature 
\cite{aharonov1993quantum, harmin1997coherent, bouwmeester1999optical, kempe2003quantum}. There each 
step is described by a unitary transformation conditioned on internal qubit states.

\subsection{Phase-space description\label{sec:phase_space}}

The theory of center-of-mass interferometry is conveniently carried out in the Wigner-Weyl
phase-space representation; its merits were repeatedly demonstrated in the context of near-field
interferometry
\cite{PhysRevA.70.053608,KDTLI,OTIMA,Varenna,PhysRevA.78.023612,NatComBateman,
dis_stef}. 
Here, we provide the phase-space counterparts of the state transformation (\ref{eq:state_trafo}).

Given a matter-wave state $\rho$ prior to the grating, the Wigner function is defined as
\begin{equation}
w(x,p)=\frac{1}{2\pi\hbar}\int d s \:  e^{ i ps/\hbar}\left\langle
x-\frac{s}{2}\right|\rho\left|x+\frac{s}{2}\right\rangle.\label{eq:wigner_def}
\end{equation}
The conditional, norm-reducing state transformation $\rho \to \mathsf{M}_{\ell} \rho 
\mathsf{M}^{\dagger}_{\ell}$ translates into a convolution in phase-space,
\begin{equation}
w(x,p)\to w (x,p;\ell) := \int d p_{0}\: w(x,p-p_{0})T_{\text{L}}(x,p_{0};\ell).
\label{eq:grating_trafo}
\end{equation}
The convolution kernel for a given absorption number $\ell$ reads
\begin{align}
T_{\text{L}}(x,p;\ell) &= \frac{1}{2\pi\hbar}\int d s\: e^{ i ps/\hbar}\: M_{\ell}\left( x -
\frac{s}{2} \right)M_{\ell}\left( x + \frac{s}{2} \right) \nonumber \\
 &=\frac{1}{2\pi\hbar}\sum_{j=-\infty}^{\infty} e^{2\pi i j x /d } \int d s\: e^{ i
ps/\hbar}B_{j}\left(\frac{s}{d};\ell\right) .\label{eq:fourier_laser}
\end{align}
The second expression is a Fourier expansion with $d = \lambda_{\mathrm{L}}/2 = 
\pi/k_{\mathrm{L}}$ the
grating period. Note that the so defined conditional Wigner function is normalized to the
absorption probability $\int d x  d p\, w(x,p;\ell) = P_{\ell} (\rho)$.

The Fourier components $B_{j}(s/d;\ell)$ generalize the Talbot coefficients, determining the
interference pattern in matter-wave interferometry \cite{KDTLI}.
For $\ell=0$ the coefficients are given by the expressions found for photo-depletion gratings 
\cite{OTIMA},
\begin{align}
B_{j}(\xi;0) =&
 e^{-n_{0}/2}\left(\frac{\zeta_{\text{coh}}-\zeta_{\text{abs}}}{\zeta_{\text{coh}}+\zeta_{\text{abs
}}}\right)^{j/2} \nonumber \\
 \times & 
J_{j}\left(\mathrm{sgn}\left(\zeta_{\text{coh}}+\zeta_{\text{abs}}\right)\sqrt{\zeta_{\text{ coh } }
^ { 2 }-\zeta_{\text{abs}}^{2}}\right). \label{eq:B_OTIMA}
\end{align}
Here, the parameters 
\begin{equation}
  \zeta_{\text{abs}}(\xi) = \frac{n_{0}}{2}\cos\left(\pi\xi\right) \textnormal{ and }
\zeta_{\text{coh}}(\xi) = \phi_{0}\sin\left(\pi\xi\right) 
\end{equation}
relate to the photon absorption and the dipole interaction, respectively.
For $\ell\neq0$ the conditional Talbot coefficients are
\begin{align}
B_{j}\left(\xi;\ell\right)
&=\sum_{n=0}^{\ell}\sum_{r=0}^{n}\left(\frac{n_{0}}{4}\right)^{n}\frac{\zeta_{\text{abs}}^{\ell-n}
(\xi)}{r!(n-r)!(\ell-n)!} \nonumber \\
&\times B_{j-n+2r}(\xi;0). \label{eq:Bm_cond}
\end{align}
When the detector is insensitive to the internal molecular state one must
resort to the unconditional state transformation, i.e.~sum over all conditional 
transformations (\ref{eq:grating_trafo}), to obtain the unconditional Wigner 
function $w'(x,p) = \sum_{\ell=0}^{\infty} w(x,p;\ell)$. This is equivalent to 
summing over the conditional Talbot coefficients in (\ref{eq:fourier_laser}), 
$B_{j}(\xi)=\sum_{\ell=0}^{\infty}B_{j}(\xi;\ell)$.
After rearranging the terms in (\ref{eq:Bm_cond}) and substituting $n$ with $m = 2r-n$ we recognize a series representation of the modified Bessel function, $I_{\nu}(z)=\sum_{k=0}^{\infty}(z/2)^{2k+\nu}/k!(\nu+k)!$. The resulting expression can be simplified further with help of Neumann's addition theorem, $\sum_{j=-\infty}^{\infty}I_{j-\nu}(u)I_{j}(v) = I_{\nu}(u+v)$, and a special case of Graf's addition theorem $\sum_{j=-\infty}^{\infty}J_{j}(u)I_{j+n}(v) = [(u-v)/(u+v)]^{n/2}J_{-n}(\mathrm{sgn}(u+v)\sqrt{u^{2}-v^{2}})$ \cite{Abramowitz}. Finally, we get the unconditional Talbot coefficients \cite{dis_stef,Cotter2015}
\begin{eqnarray}
B_{j}(\xi)&=& e^{\mathrm{-}\zeta'_{\text{abs}}}\left(\frac{\zeta_{\text{coh}}+\zeta'_{\text{abs}}}{
\zeta_{\text{coh}}-\zeta'_{\text{abs}}}\right)^{j/2} \nonumber \\
&\times&
J_{j}\left(\mathrm{sgn}\left(\zeta_{\text{coh}}-\zeta'_{\text{abs}}\right)\sqrt{\zeta_{\text{ coh }
} ^ { 2}-\left(\zeta'_{\text{abs}}\right)^{2}}\right),\label{eq:sumB}
\end{eqnarray}
with $\zeta'_{\text{abs}}=n_{0}\sin^{2}\left(\pi\xi/2\right)$.

The conditional expression (\ref{eq:Bm_cond}) applies if the molecules or nanoparticles are detected selectively according to their absorption number $\ell$. Otherwise, the expression (\ref{eq:sumB}) applies. We note that this expression resembles an earlier model \cite{KDTLI}, where photon absorption was implemented as a classical random walk in phase space, disregarding the \textit{coherent} recoil transfer in a standing wave. Surprisingly, the difference merely amounts to a sign flip $\zeta_{\text{coh}} \to - \zeta_{\text{coh}}$ in (\ref{eq:sumB}), which is equivalent to replacing $B_j (\xi)$ with $B_j (-\xi)=B_{-j} (\xi)$, as follows from the identity 
\begin{equation}
\left(\frac{y-x}{y+x}\right)^{j/2}\mathrm{sgn}(y+x)^{j}=\left(\frac{y+x}{y-x}\right)^{-j/2}\mathrm{
sgn}(y-x)^{j},
\end{equation}
where $x,y\in\mathds{R}$. Hence, the difference between the two models disappears in
the two extreme cases of no absorption, $n_0 \ll \phi_0$, and dominant absorption, $n_0 \gg \phi_0$.

\section{Absorption effects on interference}\label{sec:interference}
The results of the previous section can be readily applied to assess the effect of absorption in 
arbitrary matter-wave diffraction experiments with nanoparticles at optical standing waves. Here 
we apply our model to two 
exemplary settings: The Kapitza-Dirac-Talbot-Lau near-field matter-wave interferometer (KDTLI) and
 far-field diffraction 
at a single standing-wave grating. In both 
cases we explore the influence of absorption and evaluate the 
predicted interferograms. It turns out that the predictions of the model are fully captured only in 
the near field.

\subsection{Talbot-Lau near-field interferometer\label{sec:near_field}}
In the KDTLI setting, an initially 
incoherent beam of molecules passes through three equidistantly separated gratings
with the same period $d$; first a material grating (G1), then the standing-wave 
laser grating (G2), and finally another material grating (G3). 
Molecular matter waves emerging from each source slit at G1 obtain sufficient 
spatial coherence by propagating the distance $L$ to G2, where they are 
diffracted.
Talbot-Lau interference \cite{berman1997atom} may then yield a high-contrast 
fringe pattern of the period $d$ at a distance $L$ further downstream. G3 
serves as a movable mask to scan the interference pattern by counting the 
number of transmitted particles as a function of the lateral shift 
$x_\mathrm{s}$ of G3 relative to G1 and G2. The two material masks have the same 
opening fraction, i.e.~the same ratio $f$ between slit opening and grating 
period. 

For a detailed theoretical derivation of the predicted Talbot-Lau interference 
signal, we refer the reader to previous publications \cite{PhysRevA.78.023612, 
KDTLI, dis_stef, Varenna}. The detected signal can be expressed much like in the 
case of a coherent grating transformation by means of the Talbot 
coefficients (\ref{eq:sumB})
\begin{equation}
S\left(x_\mathrm{s}\right) =\sum_{j=-\infty}^\infty f^{2}\mathrm{sinc}^{2}\left(j\pi
f\right)B_{2j}\left(j\frac{L}{L_\mathrm{T}}\right) e^{ 2\pi i j x_\mathrm{s} /d } . 
\label{eq:signal_uncond}
\end{equation}
The Talbot length $L_\mathrm{T} = d^2 / \lambda_\mathrm{dB}$, including the 
de Broglie wavelength $\lambda_\mathrm{dB}$ of a molecule, appears as the 
natural unit for the distance between the gratings.

A robust way to quantify the fringe contrast is to fit a sine curve of period $d$ to the noisy 
measurement data. The ratio between amplitude and offset, i.e.~the sine visibility, corresponds to 
the ratio of the first and the zeroth Fourier component in (\ref{eq:signal_uncond}),
\begin{equation}
 \mathcal{V}_{\text{sin}} \left(\frac{L}{L_\mathrm{T}}\right)  = 2\mathrm{sinc^{2}}\left(\pi 
f\right) 
B_{2}\left(\frac{L}{L _\mathrm{T}}\right). \label{eq:Vl}
\end{equation} 
Negative values indicate a phase-flipped interference pattern. In the 
envisaged KDTLI setup, the sinusoidal visibility is close to the 
visibility $\mathcal{V} = (S_{\max}-S_{\min}) / (S_{\max}+S_{\min})$ defined 
in terms of the interference minima and maxima.

\begin{figure}
 \includegraphics[width=0.48\textwidth]{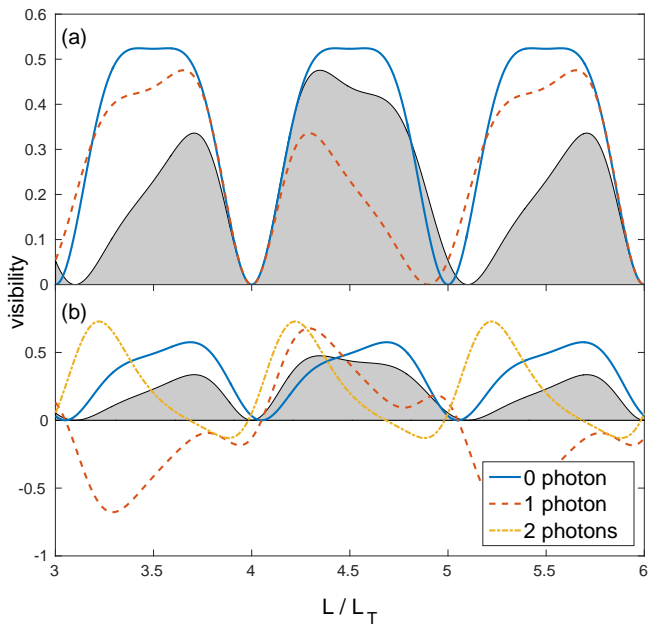} \caption{%
 (a) Sinusoidal visibility (\ref{eq:Vl}) of KDTLI as a function of the Talbot parameter 
$L/L_\mathrm{T}$, i.e.~the grating distance over Talbot length $L_{\mathrm{T}}$. The case of a pure 
phase grating (solid blue line, $\phi_0 = \pi$ at $\sigma_{\mathrm{abs}} = 0$) is compared to an 
absorptive molecule with $n_0=1$ (shaded area). A classical random-walk description of absorption 
(dashed red line) does not match the correct quantum prediction.
 (b) Visibilities (\ref{eq:Vln}) of the constituent conditional interferograms corresponding to
$\ell=0,\:1,$ and $2$ photon absorptions (solid, dashed, and dash-dotted lines, respectively). The
shaded area represents the unconditional visibility as in panel (a). The opening 
fraction is $f=0.42$ both plots.  
}
\label{fig:visibility}
\end{figure}

Figure \ref{fig:visibility}(a) compares the expected visibilities for moderate 
absorption ($n_0 = 1$, grey-shaded area) with those for no absorption ($n_0 = 
0$, solid blue line), both at $\phi_0 = \pi$. One observes that, compared to the 
case of a pure phase modulation at G2, absorption decreases the unconditional 
visibility almost everywhere. This unconditional interferogram results 
from the incoherent overlay of the conditional interferograms labeled by $\ell$; 
some of these are phase-flipped with respect to the others resulting in the 
negative visibilities depicted in Panel (b). In addition in the 
absence of photon absorptions, the visibility is periodic in the grating separation $L$ with 
period $L_\mathrm{T}$ (solid line in Fig.~\ref{fig:visibility}(a)). 
Absorption breaks this symmetry and doubles the period to $2L_\mathrm{T}$. The reason is 
that the photon absorption comes with a recoil transfer in units of \emph{half} the grating 
momentum, $\hbar
k_{\mathrm{L}}$.

Curiously, our measurement-based model for absorption predicts visibilities that
look like a mirror image of those from a classical random-walk model (red dashed line in 
Fig.~\ref{fig:visibility}(a)). This
difference was not observed in previous experiments \cite{KDTLI,NatKDTLI} because it only shows up
in interference patterns recorded with a sufficiently narrow velocity distribution. Recent
experiments with improved velocity selection \cite{Cotter2015} reveal the
model discrepancy and provide evidence for the present quantum model.

The unconditional fringe signal (\ref{eq:signal_uncond}) underlying the 
grey-shaded area in Fig.~\ref{fig:visibility}(a) and (b) is a sum of conditional 
interferograms,
\begin{equation}
S\left(x_\mathrm{s};\ell\right) = \sum_{j}f^{2}\mathrm{sinc}^{2}\left(j\pi f\right)
B_{2j}\left(j\frac{L}{L_\mathrm{T}};\ell\right) e^{ 2\pi i j x_\mathrm{s} /d}.\label{eq:signal}
\end{equation}
A molecule detector sensitive to the internal state would be able to resolve 
these interferograms. Their individual sinusoidal visibilities are given by
\begin{equation}
\mathcal{V}_{\text{sin}} \left( \frac{L}{L_\mathrm{T}};\ell \right) = 2\mathrm{sinc^{2}}\left(\pi 
f\right)\frac{B_{2}\left(L/L_\mathrm{T};\ell\right)}{B_{0}(0;\ell)}.\label{eq:Vln}
\end{equation}
They can reach as high values as 70\%, see Fig.~\ref{fig:visibility}(b).

\begin{figure}
 \includegraphics[width=0.48\textwidth]{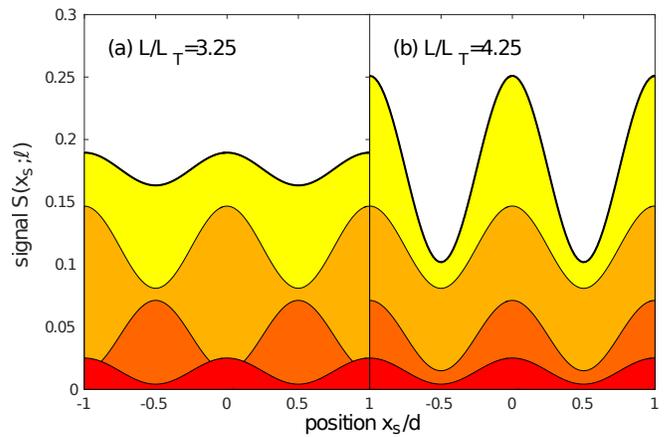} \caption{%
Panel (a) and (b) show the predicted conditional, $\ell$-dependent fringe patterns 
(\ref{eq:signal}) underlying
the visibilities in Fig.~\ref{fig:visibility}. They are plotted as a function of the lateral shift 
$x_\mathrm{s}$
of the third grating at two fixed Talbot parameters $L/L_{\mathrm{T}}=3.25$ and $4.25$. Each
panel contains the stacked conditional interferograms for molecules absorbing $\ell=0$, $1$, and 
$2$
photons (thin lines, lower mean signal for greater $\ell$-values), as well as the weighted sum over
all $\ell$ (unconditional interferogram, thick line on top). 
}
\label{fig:KDTLI_interferogram}
\end{figure}

In Fig.~\ref{fig:KDTLI_interferogram}(a) and (b) we show cascades of conditional fringe patterns
(\ref{eq:signal}) as a function of the G3 shift $x_\mathrm{s}$ for the same parameters as before 
($n_{0}=1$, 
$\phi_{0}=\pi$, $f=0.42$). The panels (a) and (b) correspond to fixed Talbot parameters $L/ 
L_\mathrm{T}=3.25$ and $4.25$, respectively. All conditional patterns (stacked thin lines) have the 
same
period $d$ as the unconditional signal (thick top line), but the odd absorption numbers can be
phase-flipped with respect to the even ones
when the Talbot parameter $L/L_\mathrm{T}$ is in the range between an odd and a next even integer. 
This is the
case in panel (a) which thus features a lower unconditional contrast than panel (b).

The relative weights of the constituent interferograms, i.e.~the transmission probabilities for 
molecules of given absorption numbers $\ell$, depend on the average absorption strength $n_0$. They 
are given by the mean value of (\ref{eq:signal}) with respect to $x_\mathrm{s}$, 
\begin{eqnarray}
\bar{S}_\ell &=& f^2 B_{0}(0;\ell)= f^2
 e^{-n_{0}/2}\left(\frac{n_{0}}{2}\right)^{\ell}
\nonumber \\
&\times&\sum_{n=0}^{\ell}\sum_{r=0}^{n}\frac{I_{2r-n}(-n_{0}/2)}{2^{n}r!(n-r)!(\ell-n)!}.
\label{eq:transmission}
\end{eqnarray}
For the case of $n_0 = 1$, illustrated in Fig.~\ref{fig:KDTLI_interferogram}, we
find that the relative weights of the conditional interferograms decrease from 64$\%$ to 24$\%$ and
8$\%$, for $\ell=0$, $1$, and $2$, respectively. The remaining 4\% of higher absorption numbers are
hardly relevant.

\subsection{Far-field interferometry}
\begin{figure}
\centering
\includegraphics[width=0.49\textwidth]{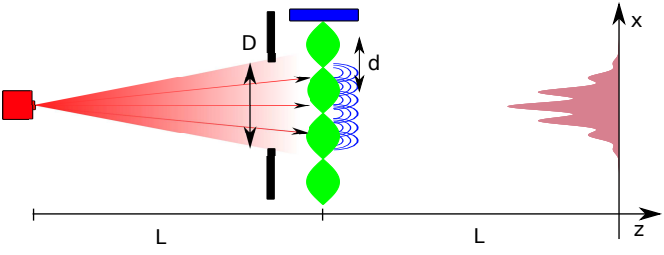}
\caption{Sketch of a symmetric far-field configuration for molecular diffraction at a standing-wave
grating. A collimated beam of molecules, as produced by an incoherent point-like source in
combination with a collimation slit of width $D$ at distance $L$, is diffracted at a standing laser
wave with grating period $d$. The resulting far-field interference pattern can be recorded by a
spatially resolving detector in distance $L$ to the laser grating.
}
\label{fig:far_setup}
\end{figure}

Let us now turn to the influence of photon absorption on 
far-field diffraction at a laser grating \cite{PhysRevLett.87.160401}. A simple 
setup is sketched in Fig.~\ref{fig:far_setup}. We consider a beam of molecules 
diverging from a point-like source and collimated by a slit aperture of width 
$D$ at distance $L$ from the source. The laser grating with period 
$d=\lambda_{\mathrm{L}}/2$ is placed immediately behind the aperture. Molecular 
matter waves are diffracted and their density distribution far from the grating 
exhibits a characteristic interference fringe pattern. For simplicity, we 
consider here a symmetric arrangement with equal distances between source, 
grating, and detection plane, as well as a monochromatic beam of molecules at 
sufficiently high forward velocity $v_z$ to allow for a one-dimensional 
phase-space treatment of the Fresnel-Kirchhoff diffraction integral in the 
paraxial approximation \cite{BornWolfBook}.

We begin with an idealized molecular point source as presented by the (unnormalized) initial Wigner 
function $w(x,p)=\delta(x)$. After free propagation by the distance $L$ described by the 
shearing transformation $w(x,p)\to w(x-pL/mv_z, p)$, the molecular beam is collimated by passing an 
aperture of width $D$. In phase space, this process is described by a convolution analogous to the 
grating transformation (\ref{eq:grating_trafo}),
\begin{eqnarray}
 w(x, p) &\to& \int_{-\infty}^{\infty} dp_0 w(x,p-p_0)\Theta\left(|x|-\frac{D}{2} \right)
\nonumber \\
& &\times  \frac{\sin[(2|x|-D)p_0/\hbar]}{\pi p_0},
\end{eqnarray}
where $\Theta$ is the Heaviside step function. The subsequent grating transformation 
(\ref{eq:grating_trafo}) followed by a further shearing transformation associated with the free 
propagation to the detector gives the conditional spatial density distribution $w(x;\ell)=\int 
dp\;w(x,p;\ell)$ on the screen,
\begin{align}
  w(x;\ell)&=\frac{d}{D\Delta x}\sum_{j=-\infty}^{\infty}\int\limits
_{-D/d}^{D/d} d q\:e^{ 2\pi i q x/\Delta x} \nonumber \\
 &\times B_{j}(q;\ell)\frac{\sin\left[
\pi\left(D/d-|q|\right)\left( j- 2q d / \Delta x\right)\right] }{ j - 2q d /
\Delta x}.\label{eq:farfield}
\end{align}
Here, $\Delta x$ denotes the distance between neighboring diffraction peaks on the screen plane,
\begin{equation}
\Delta x=\frac{h}{d}\frac{L}{mv_{z}}=d\frac{L}{L_\mathrm{T}}.
\end{equation}
Once again, the unconditional result is obtained by replacing the Talbot coefficients in 
(\ref{eq:farfield}) with (\ref{eq:sumB}).

We note that Eq.~(\ref{eq:farfield}) can be equivalently expressed as
\begin{equation}
 w(x;\ell)= \frac{d}{D\Delta x} \Bigg\vert \int_{-\infty}^\infty dq
e^{ 2\pi i q \left( x-d\:q \right)/\Delta x } t_\ell(d\:q) \Bigg\vert^2,
\label{eq:wl}
\end{equation} 
where the function $t_\ell(x)=\Theta(\vert x \vert -D/2)M_\ell(x)$ describes the conditional 
state
transformation due to a collimator and an optical grating. Equation (\ref{eq:wl}) has 
the form of a Kirchhoff integral within the Fresnel approximation \cite{BornWolfBook}.

\begin{figure}
\includegraphics[width=0.48\textwidth]{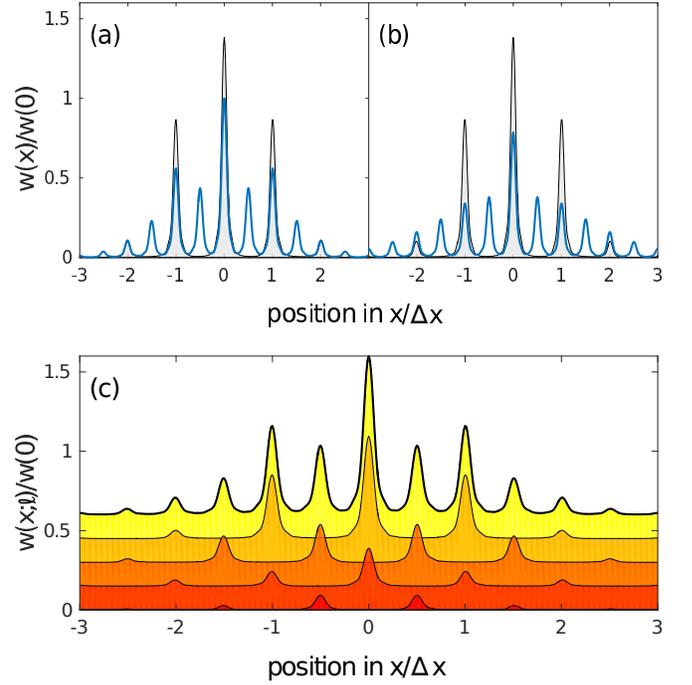} \caption{Far-field interferograms
on the screen plane behind a laser grating for non-absorbing (shaded area) and for absorbing (solid
line) molecules, assuming state-indiscriminate detection. Panel (a) and (b) correspond
to two different absorption strengths, $n_{0}=2$ and $n_{0}=10$, respectively. The shaded curves 
represent diffraction at a pure phase grating ($n_0=0$) at
$\phi_0 = 2.5$. The screen coordinate is given
in units of the expected separations $\Delta x$ of the coherent
diffraction maxima. (c) Conditional far-field interferograms contributing to the
unconditional fringe signal in Panel (a); the curves are shifted vertically for better 
illustration. The thin lines from top to bottom are the result of evaluating
(\ref{eq:farfield}) for $\ell = 0,1,2$ photon absorptions, respectively. The thick line represents 
the unconditional
result depicted in Panel (a), i.e.~the incoherent sum over all contributions. 
For this plot, we
assume a finite detector resolution of $0.1\Delta x$ and a collimator width of $D/d=10$.
}
\label{fig:far_n0}
\end{figure}

The effect of absorption on far-field interferograms is illustrated in Fig.~\ref{fig:far_n0}. We 
present the expected density distribution on the screen with and 
without absorption. In Panels (a) and (b), the shaded areas stand for a pure 
phase-grating, i.e.~for molecules with a vanishing absorption cross section ($n_0=0$), whereas the 
solid line corresponds to the unconditional interferogram of strongly absorbing molecules with $n_0 
= 2$ (left) and $n_0 = 10$ (right). In all cases the phase modulation is assumed to be $\phi_0 = 
2.5$.  
One observes that the coherent diffraction peaks at integer multiples of $\Delta x$ get reduced if 
there is a finite absorption probability, while density peaks at half integer
multiples of $\Delta x$ get populated. This is again related to the recoil momentum upon 
absorption of 
half a grating momentum, as becomes apparent in Fig.~\ref{fig:far_n0}(c), where the 
unconditional 
result
(thick line, $n_0=2$) is decomposed into its components (\ref{eq:farfield}) representing the 
conditional
interferograms for fixed absorption numbers $\ell = 0,1,2$ (thin lines from top to bottom). Odd
absorption numbers are responsible for the additional peaks as they have their diffraction peaks
only at odd halves of $\Delta x$.

We note that a classical random-walk model for absorption produces far-field interferograms that are 
almost identical to the  results plotted in Fig.~\ref{fig:far_n0}. In fact, both models give 
indistinguishable predictions for Fraunhofer diffraction. This can be seen by carrying out the 
Fraunhofer far-field approximation $d/\Delta x\ll 1$ in (\ref{eq:farfield}),
\begin{eqnarray}
w(x;\ell)&\simeq& \sum_{j=-\infty}^{\infty}\int_{-D/d}^{D/d} d q\: e^{ 2\pi i qx/\Delta
x}B_{j}(q;\ell) \nonumber \\
&\times&\frac{\sin\left[ \pi\left(D/d-|q|\right)j
\right]}{j}.\label{eq:farfield_far}
\end{eqnarray}
This expression is invariant under the sign flip $ j \to -j$, so that 
there is no difference between both models in the far-field limit, see Sect.~\ref{sec:phase_space}.
This means that the coherence in the photon momentum transfer in a standing-wave grating can only be 
observed in the near field. The KDTLI setup with sufficiently absorptive molecules, where the laser 
is neither a pure phase grating nor purely absorptive, is well suited for this purpose 
\cite{Cotter2015}.

\section{Dynamical description of the optical grating\label{sec:master}}

In this section, we present a dynamical description of the interplay between the center-of
mass-motion and the internal state evolution of a molecule interacting with a standing laser wave. 
First we introduce a ladder model for the photon absorption of molecules whose internal degrees of
freedom act as an effective heat sink. The resulting master equation for the center-of-mass state 
of the
molecule will be found to corroborate the measurement-based model for absorption given in
Sect.~\ref{sec:optical-grating}. The model is then generalized to include state-dependent internal 
properties, which is potentially relevant in experimental scenarios involving highly absorptive 
nanoparticles.

Finally, our phenomenological treatment of incoherent absorption will be compared to an 
effective three-level Rabi model for the molecule-light interaction with a finite degree of 
coherence. It shows that if Rabi oscillations occur they can have a significant impact on the 
interference pattern.

\subsection{Effective ladder model for absorbing particles}

We consider the following simple absorption model for particles that can absorb several photons 
without re-emission and whose only known properties are the polarizability and the absorption 
cross-section: Starting from a particle in its internal ground state $|0\rangle$, every subsequent 
photon absorption shall excite the internal state to distinct orthogonal states $|0\rangle \to 
|1\rangle \to |2\rangle \ldots$ of increasing internal energies $E_{\ell}=E_0+\ell\hbar \omega$. 
The total state during the interaction with the light field is then described by 
the time-dependent density matrix $\bra{x,\ell} \rho \ket{x',\ell'}$. The goal is to find 
expressions for the conditional and the unconditional states after the interaction, i.e.~for the 
projections $\rho_{\ell\ell}(x,x'; t) := \bra{x,\ell} \rho \ket{x',\ell}$ and for the reduced 
center-of-mass operator $\rho(x,x';t)=\sum_{\ell=0}^{\infty}\rho_{\ell\ell}(x,x';t)$.

In a one-dimensional description, the particle interacts with the laser while it crosses the 
Gaussian intensity profile (\ref{eq:intensity}) at a fixed longitudinal velocity $v_z$. This 
results in a mean interaction time $t_{\mathrm{L}}=\sqrt{\pi/2} w_{z}/v_{z}$ and in a time-dependent 
Hamiltonian
\begin{equation}
\mathsf{H}=\sum_{\ell = 0}^\infty \left[ E_{\ell} + V_{\ell} (\mathsf{x},t) \right] |\ell\rangle\langle\ell|.
\end{equation}
The kinetic energy term is omitted since we are neglecting the transverse motion of the particle
during its passage through the laser grating, see Sect.~\ref{sec:optical-grating}. The
$\ell$-dependence of the dipole interaction potential takes into account that the 
particle's
polarizability will in general depend on its internal state.

Photon absorption can be described as a random jump process in terms of a Lindblad-type master 
equation
\cite{PhysRevA.49.4180,wiseman2009quantum}, with the jump rate set by the time-dependent absorption
rate at the antinodes, $\gamma_{\ell}(t) = 8 \sigma_{\mathrm{abs},\ell} P / \pi w_y w_z \hbar
\omega_{\mathrm{L}}\exp(-2(v_z t)^2/\omega_z^2)$, see Eq.~(\ref{eq:n}).
In general, the absorption cross-section may depend on the internal state.
When an absorption event occurs, two effects must be considered: an excitation of the internal state up the ladder, $\ell\to\ell+1$, and the coherent
transfer of photon recoil from the standing wave to the particle.
Both effects can be implemented by introducing the Lindblad operator
\begin{equation}
\mathsf{L} = \sum_{\ell = 0}^\infty \sqrt{\gamma_{\ell} (t)} \cos \left( k_{\mathrm{L}} \mathsf{x} \right) \ket{\ell+1} \bra{\ell},
\end{equation}
which correlates the internal and external state of the particle. The evolution of the density
operator follows the master equation $ \partial_t \rho = [\mathsf{H},\rho]/i\hbar +
\mathsf{L}\rho\mathsf{L}^{\dagger} - \{ \mathsf{L}^{\dagger}\mathsf{L},\rho \}/2$. Expanded in the
basis of internal states, we are left with a sequence of coupled ordinary differential equations
that are diagonal in position representation. The internally diagonal terms $\rho_{\ell\ell}
(x,x';t)$ of interest decouple from the rest and yield a closed set of equations,
\begin{align}
\partial_t \rho_{00} &= \Bigg[ \frac{V_0(x,t) - V_0(x',t)}{ i \hbar} - \gamma_0 (t)
\nonumber \\
&\times  \frac{\cos^2 (k_{\mathrm{L}}x)+\cos^2 (k_{\mathrm{L}}x')}{2}
\Bigg]\rho_{00},\label{eq:master_d0}\\
\partial_t \rho_{\ell\ell} &= \Bigg[ \frac{V_{\ell}(x,t) - V_{\ell}(x',t)}{ i \hbar} -
\gamma_{\ell} (t)\nonumber \\
&\times \frac{\cos^2 (k_{\mathrm{L}}x)+\cos^2 (k_{\mathrm{L}}x')}{2}
\Bigg]\rho_{\ell\ell} +\gamma_{\ell-1} (t)  \nonumber
\\
 &\times \cos (k_{\mathrm{L}}x)\cos (k_{\mathrm{L}}x')\rho_{\ell-1\ell-1}.  \label{eq:master_dl} 
\end{align}
The imaginary terms represent the coherent phase modulation due to the dipole interaction, while the
other terms describe the redistribution of internal state populations according to the rates
$\gamma_{\ell} (t)$. The redistribution is Poissonian if all rates are equal. The first equation can
be integrated directly, and the remaining sequence of equations can then be solved successively
starting from the initial condition $\rho_{\ell\ell} (x,x';-\infty) = \widetilde{\rho} (x,x')
\delta_{\ell,0}$.

In the simple case of a state-independent dipole potential $V(x,t)$ and absorption rate $\gamma_0(t)$, the outgoing solution can be written in compact form,
\begin{equation}
\rho_{\ell\ell}(x,x',\infty)= 
M_{\ell}(x)M_{\ell}(x')\widetilde{\rho}(x,x').\label{eq:rho_ell}
\end{equation}
This reproduces our measurement-based result $M_{\ell} (x)$
defined as (\ref{eq:measurment}) and $n_0 = \int  d t\, \gamma_0(t)$.

\subsection{Generalized model \label{sec:effect_of_excited}}

We proceed to generalize the ladder model to the case where the molecular parameters switch to a 
fixed excited-state value after the absorption of at least one photon. For this, we introduce 
dimensionless parameters $\eta_{\mathrm{p}}$ and $\eta_{\mathrm{a}}$ describing the changed 
excited-state dipole potential and absorption rate, $V_{\ell}=\eta_{\mathrm{p}}V_{0}$
and $\gamma_{\ell}=\eta_{\mathrm{a}}\gamma_{0}$ for all $\ell>0$.
Models with a stepwise increase of the polarizability and absorption rate have been employed for 
the determination of molecular
excited-state properties \cite{Henari1992144,Gratz199921}. 

For simplicity, we approximate the Gaussian laser envelope by a constant intensity switched on for the effective interaction time $t_{\mathrm{L}}=\sqrt{\pi/2}\omega_{z}/v_{z}$. This assumption, which leaves the time-integrated parameters $\phi_0$ and $n_0$ of the Poisson model unchanged, is well justified in the Raman-Nath regime and produces analytical results. The coupled equations (\ref{eq:master_d0})
and (\ref{eq:master_dl}) for $\ell>0$ can be evaluated to
\begin{align}
\rho_{\ell\ell}(x,x',t_\mathrm{L}) &= \widetilde{\rho}(x,x^\prime) M_{\ell}(x)M_{\ell}(x')\nonumber 
\\
&\times \eta_{\mathrm{a}}^{\ell-1}{}_{1}F_{1}(\ell;\ell+1;z(x,x')), \label{eq:rho_n_eta} 
\end{align}
where we introduced the abbreviation
\begin{equation} 
z(x,x') =  i
(\eta_{\mathrm{p}}-1)[\phi(x)-\phi(x')]-\frac{\eta_{\mathrm{a}}-1}{2} [n(x)+n(x^\prime)] \\
\end{equation}
\begin{figure}
\includegraphics[width=0.5\textwidth]{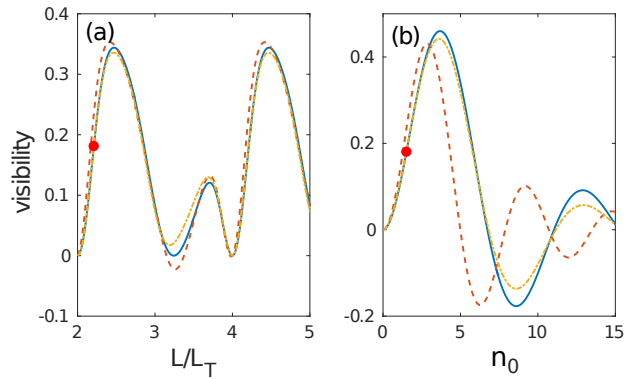}\caption{(a) Sinusoidal visibility as
a function of the Talbot parameter $L/L_{\mathrm{T}}$ for molecules with different excited-state
parameters. For the solid curve, we assume that the ground-state values of polarizability and
absorption cross-section (corresponding to $n_{0}=1.5$ and $\phi_{0}=1.25n_{0}$) remain the same no
matter how many photons are absorbed. The dashed and the dash-dotted curve correspond
to an increase of the absorption cross-section and of the polarizability respectively by a factor 
of $1.5$ upon absorption of
the first photon. 
(b) Same visibilities as a function of laser power, i.e.~for varying $n_0$ at a fixed Talbot
parameter of $L/L_{\mathrm{T}}=2.2$. For reference, the red dot marks the same spot 
$(L/L_{\mathrm{T}},n_0)$ in both panels. 
A significant difference between the curves appears at high laser powers.}
\label{fig:vis_eta}
\end{figure}
Here, $_{1}F_{1}$ denotes the confluent hypergeometric function \cite{Abramowitz}. The solution for
$\ell=0$ is identical to the one given in (\ref{eq:rho_ell}).

It is instructive to make use of an integral representation of the hypergeometric function,
$_{1}F_{1}(\ell;\ell+1;z)=\ell\int_{0}^{1}d\alpha\:e^{z\alpha}\alpha^{\ell-1}$, to
represent (\ref{eq:rho_n_eta}) as a conditional post-measurement state,
\begin{equation}
\rho_{\ell\ell}=\frac{1}{t_{\mathrm{L}}}\int_{0}^{t_{\mathrm{L}}}dt_{1}\:
\widetilde{\mathsf{M}}_\ell(t_1) \widetilde{\rho}
\widetilde{\mathsf{M}}^\dagger_\ell(t_1).\label{eq:rho_n_eta_integral}
\end{equation}

This allows us to identify generalized measurement
operators analogous to (\ref{eq:unitary}) and (\ref{eq:Ml}), 
\begin{align}
 \widetilde{\mathsf{M}}_\ell(t_1) &=
\sqrt{\frac{(\eta_{\mathrm{a}}(1-t_{1}/t_{\mathrm{L}}))^{\ell-1}n_{0}^{\ell}}{(\ell-1)!}}\cos^{\ell}
(k_{\mathrm{L}}\mathsf{x}) \nonumber \\
&\times \exp\left\{ \left( i\phi(\mathsf{x})-\frac{n(\mathsf{x})}{2}
\right)\frac{t_1}{t_\mathrm{L}} \right\}  \nonumber \\
                              &\times \exp\left\{ \left(
i\eta_\mathrm{p}\phi(\mathsf{x})-\eta_\mathrm{a}\frac{n(\mathsf{x})}{2}
\right)\left(1-\frac{t_1}{t_\mathrm{L}}\right) \right\}.
\end{align}
They depend on a new parameter $t_{1}\in[0,t_{\mathrm{L}}]$, which can be interpreted as the time of
the first photon absorption. 
For $t_{1}=t_{\mathrm{L}}$, the operators reduce to their Poissonian counterparts of before.

Figure \ref{fig:vis_eta} compares the unconditional visibility (\ref{eq:Vl}) of the Poisson model (solid line)
to hypothetical cases where the excited-state absorption cross 
section
(dash-dotted line) or the excited-state polarizability (dashed line) are $50\%$ higher than the 
ground state values. 
It turns out that an increased excited-state cross section hardly affects the visibility even 
though it strongly affects the conditional transmission probability through the grating. 
The reason is that high-contrast interference is mainly produced by the 
light-induced phase modulation,
and the visibility is therefore more sensitive to absorption-induced changes of the molecular
polarizability. Indeed, we observe a substantial influence of an increased excited state 
polarizability at high laser power.
This might open up a novel spectroscopic application of the KDTLI scheme.

\subsection{Rabi model for partially coherent absorption\label{sec:rabi}}

So far, we have treated photon absorption by molecules incoherently, presuming that electronic 
transitions are not driven coherently by the light field due to the presence of rapid internal 
decay channels that involve the excitation of numerous rovibrational degrees of freedom. While this 
is a good approximation for many large molecules and nanoparticles, one could think of experimental 
situations
\cite{PhysRevA.32.1451,Shimoda197609} that would permit a few coherent Rabi cycles or even the use 
of Raman transitions, as in atomic beam manipulation 
\cite{PhysRevLett.112.203002,PhysRevD.80.016002}.

\begin{figure}
\centering
 \includegraphics[width=0.3\textwidth]{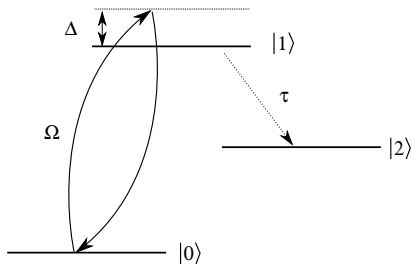}
 \caption{Scheme of a three-level Rabi model, where the laser drives the transition between the
ground state $\ket{0}$ and the excited state $\ket{1}$ at a detuning $\Delta$ off resonance. The
excited state with life time $\tau$ can decay without emission into a dark metastable state
$\ket{2}$.}
\label{fig:rabi}
\end{figure}

To study the transition from a coherent atom-like description to our model for absorption, we consider the three-level system sketched in Fig.~\ref{fig:rabi}. 
It consists of a ground state $\ket{0}$, an exited state $\ket{1}$ with the life time $\tau$, and a 
metastable dark state $\ket{2}$. The excited state shall decay exclusively to the dark state 
without emitting a photon.
This Rabi toy model is useful as it can be treated analytically. 
It is also employed in atomic experiments, e.g.~to describe absorptive optical masks \cite{abfalterer1997,andrey2003}. 
We consider the experimental situation where only ground-state molecules are detected in the end. 

We resort once again to a one-dimensional description in the Raman-Nath regime where the transverse 
motion during the interaction is negligible.
The laser shall drive the transition between $\ket{0}$ and $\ket{1}$ with a detuning $\Delta$ 
relative to the energy difference. The laser-molecule interaction is then characterized by a 
position-dependent Rabi frequency in the rotating wave approximation \cite{GerryKnight},
\begin{equation}
\Omega(x)=-\frac{\mathbf{d}_{01}\cdot \mathbf{E}_{0}}{\hbar}\cos(k_{\mathrm{L}}x) =: 
\Omega_{0}\cos(k_{\mathrm{L}}x),
\end{equation}
given the transition dipole moment $\mathbf{d}_{01}$. In a frame rotating at the laser frequency 
$\omega_{\mathrm{L}}$, the interaction Hamiltonian then reads as 
$\mathsf{H}=\frac{\hbar}{2}\left(\Omega(\mathsf{x})\ket{1}\bra{0} - \Delta \ket{1} 
\bra{1} \right)+\text{h.c}$. For the spontaneous decay to the dark state, we introduce the 
jump 
operator $\mathsf{L}=\ket{2}\bra{1} / \sqrt{\tau}$, which yields the final master equation $ 
\partial_t \rho = [\mathsf{H},\rho]/i\hbar + \mathsf{L}\rho\mathsf{L}^{\dagger} - \{ 
\mathsf{L}^{\dagger}\mathsf{L},\rho \}/2$. Note that a radiative decay would imply a more 
complicated decoherence master equation \cite{JModOptKnight}.

\begin{figure*}
\includegraphics[width=\textwidth]{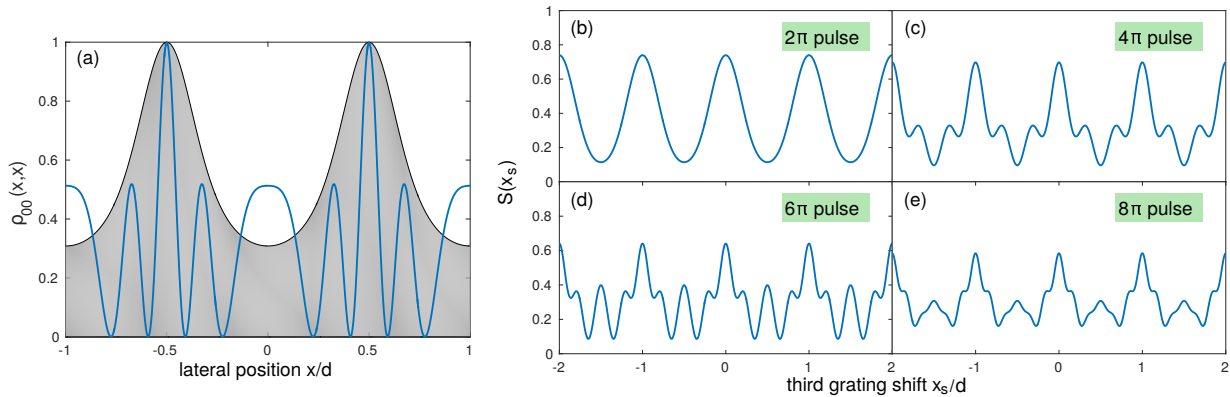}
\caption{(a) Probability density for the transmission of ground-state molecules through a standing
wave of period $d = \lambda_{\mathrm{L}}/2$ as a function of position. The shaded area corresponds
to the conditional probability for zero absorptions, taken from the ladder model
(\ref{eq:transmission}) with $n_{0}=1.2$. An evaluation of the coherent three-level Rabi model
yields the blue solid line, assuming $\Omega_{0}t_{\mathrm{L}}=4\pi$ (i.e.~a $4\pi$ pulse length at
the antinodes) and a long excited-state life time of $\tau = t_{\mathrm{L}}$. (b)-(e) Numerical
results for the KDTLI fringe signal 
in the presence of molecular Rabi oscillations, according to the three-level model on resonance
($\Delta=0$, $\tau = t_{\mathrm{L}}$). We use a setup with $L = 2L_{T}$ grating separation, and
$f=0.1$ opening fraction for G1 and G3; only ground-state molecules are detected. The antinode
intensity is chosen such that it amounts to an effective pulse length $\Omega_0 t_{\mathrm{L}}$
increasing from $2\pi$ to $8\pi$ in panels (b) to (e), respectively. We observe higher-order fringes
emerging with each Rabi cycle. }
\label{fig:KDTLIrabi}
\end{figure*}

A decomposition of the density operator into the matrix elements $\rho_{nn'} 
(x,x';t):=\bra{x,n}\rho\ket{x',n'}$ yields a linear system of nine partially coupled differential 
equations. The system can be diagonalized exactly, but we will omit the lengthy general solutions 
and focus on limiting cases.

In the limit of no decay, $\tau \to \infty$, we obtain the well-known Rabi oscillation between 
ground and excited state at the position-dependent frequency 
$\Omega_{R}(x)=\sqrt{\Delta^{2}+\Omega^{2}(x)}$ \cite{GerryKnight}. For a finite decay time 
comparable to the interaction period, the coherences $\rho_{01}$ and $\rho_{10}$ get exponentially 
suppressed,
which leads to a damping of the Rabi oscillations and to a population transfer to the dark state.

In contrast, if the excitation life time is short, $\tau \ll t_{\mathrm{L}}$, the oscillation dies 
out before a Rabi cycle is completed.
Dropping all terms containing the fast damping $\exp(-t_\mathrm{L}/\tau)$ and expanding the 
eigenfrequencies of the system to lowest order in $\tau / t_{\mathrm{L}}$, the approximate matrix 
element of the ground-state density operator reads
\begin{align}
& \rho_{00}(x,x') \simeq  \nonumber \\  
& \exp\left\{ -\frac{1}{2}\frac{t_{\mathrm{L}}
\tau\Omega_{0}^{2}}{1+4\Delta^{2}\tau^{2}}\left[ \cos^{2}(k_{\mathrm{L}}x)+\cos^{2}(k_{\mathrm{L}}
x')\right ]\right\} \nonumber \\
 & \times\exp\left\{ - i \frac{t_{\mathrm{L}}
\Delta\tau^{2}\Omega_{0}^{2}}{1+4\Delta^{2}\tau^{2}} 
\left[ \cos^{2}(k_{\mathrm{L}}x)-\cos^{2}(k_{\mathrm{L
}}x') \right] \right\} .\label{eq:rabi_rho11}
\end{align}
We notice that this coincides with the conditional density matrix (\ref{eq:rho_ell}) for zero absorptions from the above ladder model. That is, the molecule acts as an incoherent 1-photon absorber in this limit, and we can identify the effective phase shift and mean absorption number parameters by comparison, 
\begin{equation}
\phi_{0}\hat{=}-\frac{t_{\mathrm{L}} \Delta\tau^{2}\Omega_{0}^{2}}{1+4\Delta^{2}\tau^{2}},\hspace{1em}n_{0}\hat{=}\frac{t_{\mathrm{L}} \tau\Omega_{0}^{2}}{1+4\Delta^{2}\tau^{2}}.\label{eq:rabi_phi0}
\end{equation}
Similar results were derived and discussed in the case of Bragg diffraction \cite{LaserPhys.1.1}. 
As demonstrated in atomic experiments, a pure phase grating can be realized in the far off-resonant
case \cite{PhysRevLett.75.2633}, with $\phi_{0} \approx -t_{\mathrm{L}} \Omega_{0}^{2}/4\Delta$, and
a pure absorptive grating in the resonant case \cite{abfalterer1997,andrey2003}, with $n_{0} =
t_{\mathrm{L}} \tau \Omega_{0}^{2}$ 
\footnote{An additional prefactor of $4\sqrt{\pi/2}$ appears in the references for $\phi_{0}$ and
$n_{0}$, because of a different definition of the effective electric field amplitude.}.

When the interaction time is comparable to the decay time of the excited
state, the time evolution is governed by several cycles of damped Rabi oscillations, and the 
transmission of ground-state molecules will depend in an oscillatory fashion on the precise value 
$t_{\mathrm{L}}$ of the interaction time. We illustrate this in Fig.~\ref{fig:KDTLIrabi}(a), which 
depicts the position-dependent probability (blue solid line) that a molecule passes a resonant 
standing laser wave in the ground state. We assume an interaction of $\Omega_0 t_{\mathrm{L}} = 
4\pi$ mimicking a $4\pi$ pulse at the antinodes, and $\tau = t_{\mathrm{L}}$. This is compared to 
the conditional transmission probability for zero absorptions in the ladder model using $n_0 = 1.2$ 
(shaded area). 
For the Rabi case, one can observe an oscillation of the transmission probability. 
Here the minima correspond to $\pi$-pulses \cite{GerryKnight}, where all particles are 
either in the excited or in the dark state.
At an antinode, the transmission is below $100\%$ due to losses 
into the dark state.

As a consequence of the Rabi features in the laser grating, higher-order fringe oscillations should 
appear in the molecular near-field interferograms. This is demonstrated for KDTLI in the four 
interferograms of Fig.~\ref{fig:KDTLIrabi}(b)-(e). They were computed by evaluating the ground-state 
density operator $\rho_{00} (x,x';t_{\mathrm{L}})$ and inserting this solution into expression 
(\ref{eq:signal}).

This assumes that excited-state molecules decay to the dark state before detection, and that dark-state molecules are not recorded. As before, we consider the resonant situation, $\Delta =0$, and a fairly long life time, $\tau = t_{\mathrm{L}}$. The parameters were chosen such that the light intensity at the 
antinodes and the interaction time amount to an effective pulse length of one to four full Rabi 
cycles in panel (b)-(e). Here, we assume the KDTL setup to operate at a grating separation of two 
Talbot lengths and with a small open fraction $f=0.1$ at G1 and G3, not to wash out higher 
harmonics in the fringe pattern. 
The panels clearly show the appearance of these higher harmonics emerging with each additional Rabi 
cycle during the interaction time.

Higher fringe oscillations may serve to increase the phase sensitivity of near-field interference schemes with standing-wave gratings, thus boosting the precision in potential metrological applications.

\section{Conclusions and outlook}

We have presented a measurement-based model for photon absorption at standing laser waves to 
describe matter-wave diffraction at laser gratings. The model is particularly well suited for 
complex molecules and clusters which can dissipate the heat of several light quanta amongst their 
numerous internal degrees of freedom.  We noted a subtle and intricate difference compared to a 
classical random-walk model for absorption, which goes unnoticed in far-field diffraction. In the 
near field it can be observed that quantum 
interference prevails even in the case of significant absorption and state-insensitive particle 
detection -- a consequence of the interplay between coherent phase modulation at the 
standing-wave potential and a discrete coherent random walk in steps of single photon recoils in 
momentum space. Only recently, measurements in a near-field KDTLI setup with $\mathrm{C}_{70}$ 
molecules provided sound experimental evidence for the validity of our measurement-based model 
\cite{Cotter2015}, which is also corroborated by a dynamical master-equation approach based solely 
on phenomenological parameters:
The dipole polarizability and the absorption cross-section of the particle. We also showed that 
our approach is extendable to the more general case of parameters that depend on the internal
state of the particle.

Finally, we studied the impact of coherent Rabi cycles on the absorption behavior of molecules
in cases where the photo-induced internal excitation has a sufficiently long life time.
We found that Rabi oscillations imprint an additional oscillatory structure onto 
the particle state upon transmission through a laser grating, which creates higher 
harmonics in near-field interferograms. This
may be relevant for increasing the precision in potential metrological applications.

\begin{acknowledgments}
We thank M. Arndt, J. P. Cotter, S. Eibenberger, and L. Maierhofer for helpful discussions 
and the fruitful collaboration. We acknowledge support from the European Commission within 
NANOQUESTFIT (Contract No. 304886). 
S.N. is supported by the National Research Foundation, Prime Minister's Office, Singapore 
and the Ministry of Education, Singapore under the Research Centres of Excellence programme.
\end{acknowledgments}

%

\end{document}